\documentclass[aps, prl, amsmath,amssymb, reprint, showpacs,superscriptaddress,groupedaddress, nofootinbib ]{revtex4-1}

\usepackage{nicefrac}
\usepackage{float}
\usepackage{graphicx}
\usepackage{dcolumn}
\usepackage{bm}
\usepackage{comment}
\usepackage{soul}
\usepackage{hyperref}
\hypersetup{
	colorlinks=true,
	citecolor=blue,
	linkcolor=blue,
	filecolor=blue,      
	urlcolor=blue,
}

\begin{document}
	
\title{Structural, electronic, magnetic and transport properties of equiatomic quaternary Heusler Alloy CoRhMnGe: Theory and Experiment}

\author{Deepika Rani$^{1,\dagger}$, Enamullah$^{1,\dagger}$, K. G. Suresh$^1$, A. K. Yadav$^2$, S. N. Jha$^2$, D. Bhattacharyya$^2$, Manoj Raama Varma$^3$ and Aftab Alam$^{1,*}$}
\affiliation{$^1$Department of Physics, Indian Institute of Technology Bombay, Powai, Mumbai 400076, Maharashtra, India \\ $^2$Atomic and Molecular Physics Division, Bhabha Atomic Research centre, Mumbai-400094, India \\ $^3$National Institute for Interdisciplinary Sciences and Technology(CSIR), Thiruvananthpuram, India }
\email{aftab@iitb.ac.in}
\date{\today}

\let\thefootnote\relax\footnote{$^\dagger$ These authors have contributed equally to this work}

\begin{abstract}
	In this work, we present structural, electronic, magnetic, mechanical and transport properties of equiatomic quaternary Heusler alloy, CoRhMnGe using both theoretical and experimental techniques. A detailed structural analysis is performed using X-ray diffraction(XRD) and extended X-ray absorption fine structure(EXAFS) spectroscopy. The alloy is found to crystallize in Y-type structure having space group $F\bar{4}3m$ (\# 216). The ab-initio simulation pedict half-metallic ferromagnetic characteristics leading to large spin polarization. The calculated magnetization is found to be in fair agreement with experiment as well as those predicted by the Slater-Pauling rule, which is a prerequisite for half-metallicity. The magnetic transition temperature($\mathrm{T_{C}}$) is found to be $\sim 760$ K. Measured electrical resistivity in the temperature range 2-400 K also gives an indication of half-metallic behavior. Simulated resistivity matches fairly well with those measured, with the temperature dependant carrier relaxation time lying in the range $1-2$ fs. Effect of hydrostatic pressure on electronic structure, magnetic and mechanical properties are investigated in detail. The alloy is found to preserve half-metallic characteristics upto 30.27 GPa beyond which it transit to metallic phase. No magnetic phase transition is found to occur in the whole range of pressure. The system also satisfies the Born-Huang criteria for mechanical stability upto a limited range of pressure. All these properties make CoRhMnGe alloy promising for spintronics devices.
\end{abstract}
\maketitle

\section{Introduction} In the last few years, spintronics has emerged as a new field of research, utilizing spin currents, instead of electrons, as one of mediator for transport. It has proved to be quite promising for new types of fast electronic devices over conventional electrical charge-based semiconductor devices with advantages of high storage density, increased data processing speed and low power consumption.\cite{Graf2} To enhance the efficiency of spintronic devices, the carrier for the concerned materials should be fully spin polarized. Half-metallic ferromagnets (HMFs) are great source for highly spin polarized current and thus are ideal materials for spintronic based applications. Half-metallicity is found in magnetic materials in which one of the spin-polarized sub band has a band gap and the other exchange-split sub band has a non zero density of states at the Fermi level(E$_{F}$), thus in these materials the electrical conduction takes place from one spin channel exclusively. The discovery of half-metallicity in half-Heusler alloy NiMnSb,\cite{PhysRevLett.50.2024} by de Groot et al. in 1983 has intensified the interest in Heusler alloys for applications in the field of spintronics. Heusler family of compounds are usually of two types (i) a ternary material with stoichiometry 1:1:1 known as half-Heusler compounds (XYZ) (ii) also a ternary materials but with stoichiometry 2:1:1 known as full-Heusler compounds ($\mathrm{X_2YZ}$). There is another class of Heusler compounds which are discovered recently with stoichiometry 1:1:1:1, and are known as equiatomic quaternary Heusler alloys (XX'YZ), where X, X' and Y are transition metals and Z is a main group element.\cite{:/content/aip/journal/apr2/3/3/10.1063/1.4959093} The half-Heusler alloys crystallize in $C1_b$ structure (space group \#216, $F\bar{4}3m$) with prototype MgAgAs and the full Heusler alloys crystallize in the cubic space group $Fm\bar{3}m$ (\#225), with $\mathrm{Cu_2 MnAl}$ ($L2_1$) as prototype, whereas the “quaternary”  Heusler alloys crystallize in cubic space group (\#216) with LiMgPdSn as prototype (Y-type).\cite{Graf20111} Among the full Heusler compounds, considerable attention has been paid to the Co based compounds because of their high spin polarization and high Curie temperature,{\cite{Graf2, :/content/aip/journal/jap/116/20/10.1063/1.4902831,PhysRevB.91.104408, :/content/aip/journal/jap/102/3/10.1063/1.2767229,:/content/aip/journal/jap/101/2/10.1063/1.2409775}} which make them more suitable for applications in spintronics. According to Julli\`{e}re model,\cite {JULLIERE1975225} the high spin polarization is very advantageous for getting high tunneling magneto-resistance ratios in magnetic tunnel junctions. The quaternary Heusler alloys CoRhMnZ (Z = Ga, Sn and Sb) have been studied in detail by both theoretical and experimental methods and were found to be HMF by ab-initio calculations.\cite{0953} CoRhMnGe, however, has not been studied from experimental front. Hence keeping in view the increased interest and applications of Co-based Heusler alloys, we present a comprehensive study of electronic, structural, magnetic and transport properties of CoRhMnGe (CRMG) Heusler alloy. 

CRMG alloy is found to exist in the ordered cubic Heusler structure (Y-type) with space group $F\bar{4}3m$ (\#216). We have done a detailed local structure analysis using extended X-ray absorption fine structure (EXAFS) spectroscopy. The saturation magnetization value $M_s$ at 5 K is found to be 4.9 $\mu_{B}/f.u.$ which is in close agreement with the value predicted by Slater-Pauling rule (5 $\mu_{B}/f.u.$) for half-metallic materials. The Curie temperature ($\mathrm{T_{C}}$) was found to be $\sim 760$ K which is highest among the reported CoRhMnZ (Z = Ga, Sn, Sb) alloy.\cite {0953} We have also performed a systematic ab-initio calculations to study the electronic structure, magnetism, mechanical and transport properties of the alloy. Ab-initio simulations also predict half-metallic nature for this alloy. Total energy and lattice dynamics calculations suggest that the alloy is chemically and mechanically stable against external pressure. Effect of pressure on magnetism and half metallicity is studied at length.

\section{Experimental Details}

\subsection{Sample Synthesis}
The polycrystalline alloy CRMG was prepared by arc melting the stoichiometric amounts of constituent elements (at least 99.9\% purity) in water cooled copper hearth under high purity argon atmosphere. To further reduce the contamination a Ti ingot was used as an oxygen getter. 2\%  extra Mn was taken to compensate the weight loss due to Mn evaporation during melting. The ingot formed was flipped and melted several times for better homogeneity, The final weight loss was less than 1\%. 

\subsection{Characterization}
X-ray diffraction (XRD) pattern was taken at room temperature using X”pert pro diffractometer with $\mathrm{Cu-K\alpha}$ radiation to study the crystal structure of the sample. XRD analysis was done with the help of FullProf suite which uses the least square refinement between the experimental and calculated intensities. Rietveld method is used to optimize the $\chi$-square function given by:
\begin{equation}
\chi^2= w_i\Sigma_i{(y_{iobs}-y_{ical})^2}
\end{equation}
where $w_i$ is the inverse of the variance associated with the $i^{th}$ observation i.e $\sigma^2(y_{iobs})$ and $y_{iobs}$ and $y_{ical}$ are the observed and calculated scattering intensities for a diffraction angle $2\theta_i$.\cite{RR}
It also contains GFourier program, which is used to calculate and visualize the electron density within the unit cell. 

EXAFS (Extended X-ray Absorption Fine Structure) measurements on CRMG were done to probe the local structure surrounding the Co, Ge and Mn sites. The X-ray absorption spectroscopy (XAS) measurements have been carried out at the Energy-Scanning EXAFS beam-line (BL-9) in transmission mode at the INDUS-2 Synchrotron Source (2.5 GeV, 200 mA) at Raja Ramanna Centre for Advanced Technology (RRCAT), Indore, India.\cite{2014AIPC.1591..649P,1742-6596-493-1-012032} This beam-line operates in the energy range of 4-25 keV. The beam-line optics consists of a Rh/Pt coated collimating meridional cylindrical mirror and the collimated beam reflected by the mirror is monochromatized by a Si(111) (2d=6.2709 \AA ) based double crystal monochromator (DCM). The second crystal of the DCM is a sagittal cylinder used for horizontal focusing while a Rh/Pt coated bendable post mirror facing down is used for vertical focusing of the beam at the sample position. Three ionization chambers (300 mm length each) were used for data collection in the transmission mode; one for measuring incident flux ($I_0$), one for measuring transmitted flux ($I_T$) and the third for measuring EXAFS spectrum of a reference metal foil for energy calibration. Appropriate gas pressure and gas mixture have been chosen to achieve 10-20\% absorption in first ionization chamber and 70-90\% absorption in second one to improve the signal to noise ratio. Rejection of the higher harmonics content in the X-ray beam was  performed by the second mirror. The absorption coefficient, $\mu$ was obtained using the relation: 
\begin{equation}
I_T = I_0 e^{-\mu x}
\end{equation}
where, $x$ is the thickness of the absorber. Powder samples of appropriate weight, estimated to obtain a reasonable edge jump, were taken and was mixed thoroughly with cellulose powder to obtain total weight of 100 mg. Subsequently, homogeneous pellets of $15$ mm diameter were prepared using an electrically operated hydraulic press. However, the grain size of the particles was of the order of 50 micron, which resulted in very bad data. The sample was later grounded continuously using a mortar-pestle for 3 hours to reduce the particle size to less than 4 microns. This fine powder was dispersed on the scotch tape and larger particles are brushed out. A reasonable edge jump was obtained by folding the scotch tape.

Magnetization isotherms at 5 K and 300 K were obtained using a vibrating sample magneto meter (VSM) attached to the physical property measurement system (PPMS) (Quantum design) for fields up to 50 kOe. Thermo-magnetic curves in the high temperature region (300 K -1000 K) were taken in VSM with attached high temperature oven under a field of 100 Oe. Electrical resistivity measurements were done using four-probe method in PPMS.

\section{Computational Details}
Ab-initio calculations were performed to study the electronic structure, magnetic and mechanical properties using density functional theory (DFT) implemented within Vienna ab-initio simulation package (VASP)\cite{VASP} with a projected augmented wave (PAW) basis.\cite{PAW} The electronic exchange-correlation potential due to Perdew-Bueke-Ernzerhof (PBE) is used within generalized gradient approximation (GGA) scheme. A $24^3$ {\bf k}-mesh is used to perform the Brillouin zone integration within the tetrahedron method. A plane wave energy cut-off of $269$ eV is used for all the calculations. ALL the structure are fully relaxed, with total energies (forces) converged to values less than 10$^{-6}$ eV ($0.01$ eV/\r{A}).

Transport properties are calculated using Boltzmann transport theory implemented within the BoltzTrap code\cite{Madsen200667} under constant relaxation time approximation of the charge carriers. These calculations are performed using experimental lattice parameter to make a direct comparison with the measured resistivity. 
Simulating accurate transport properties usually require a much finer k-mesh as well as plane-wave energy cut-off. As such $46^3$ k-mesh along with a plane wave cut-off of 500 eV are used for transport study. Because, simulated electrical conductivity within BoltzTrap are more reliable at relatively higher temperature ($> 100$ K), we compared our calculated results with those of experiment between the range $100-400$ K.

\begin{figure}[t]
	\centering
	\includegraphics[scale=0.75]{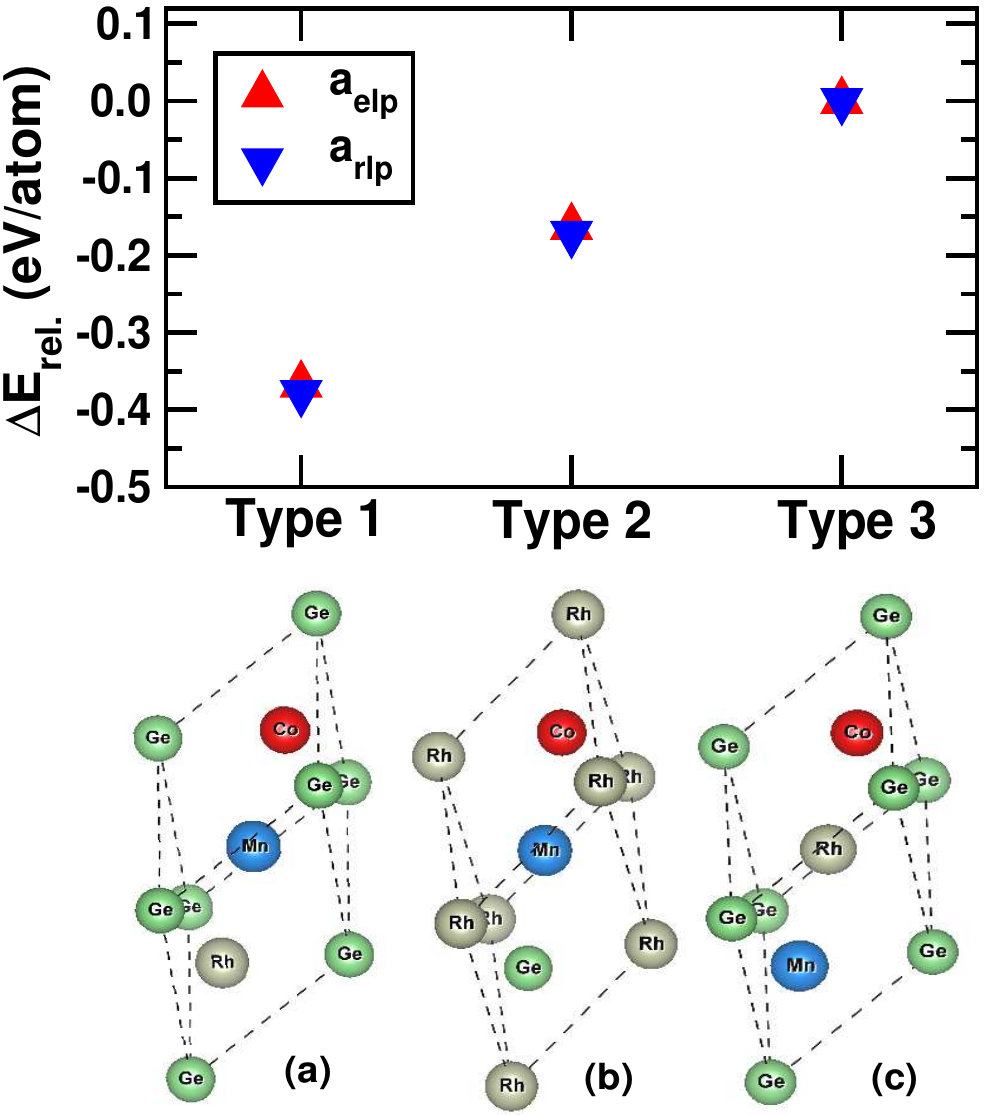}
	\caption{Relative energy ($\Delta$E$_{rel.}$) of different non-degenerate atomic configurations with respect to Type3 configuration, calculated at experimental (a$_{elp}$) and relaxed (a$_{rlp}$) lattice parameters. Three distinct configurations, Type1, Type2 and Type3 are shown by the primitive unit cell (a), (b) and (c) respectively and are detailed in the text. }
	\label{fig1}
\end{figure}

\begin{figure}[t]
	\centering
	\includegraphics[width=1.0\linewidth]{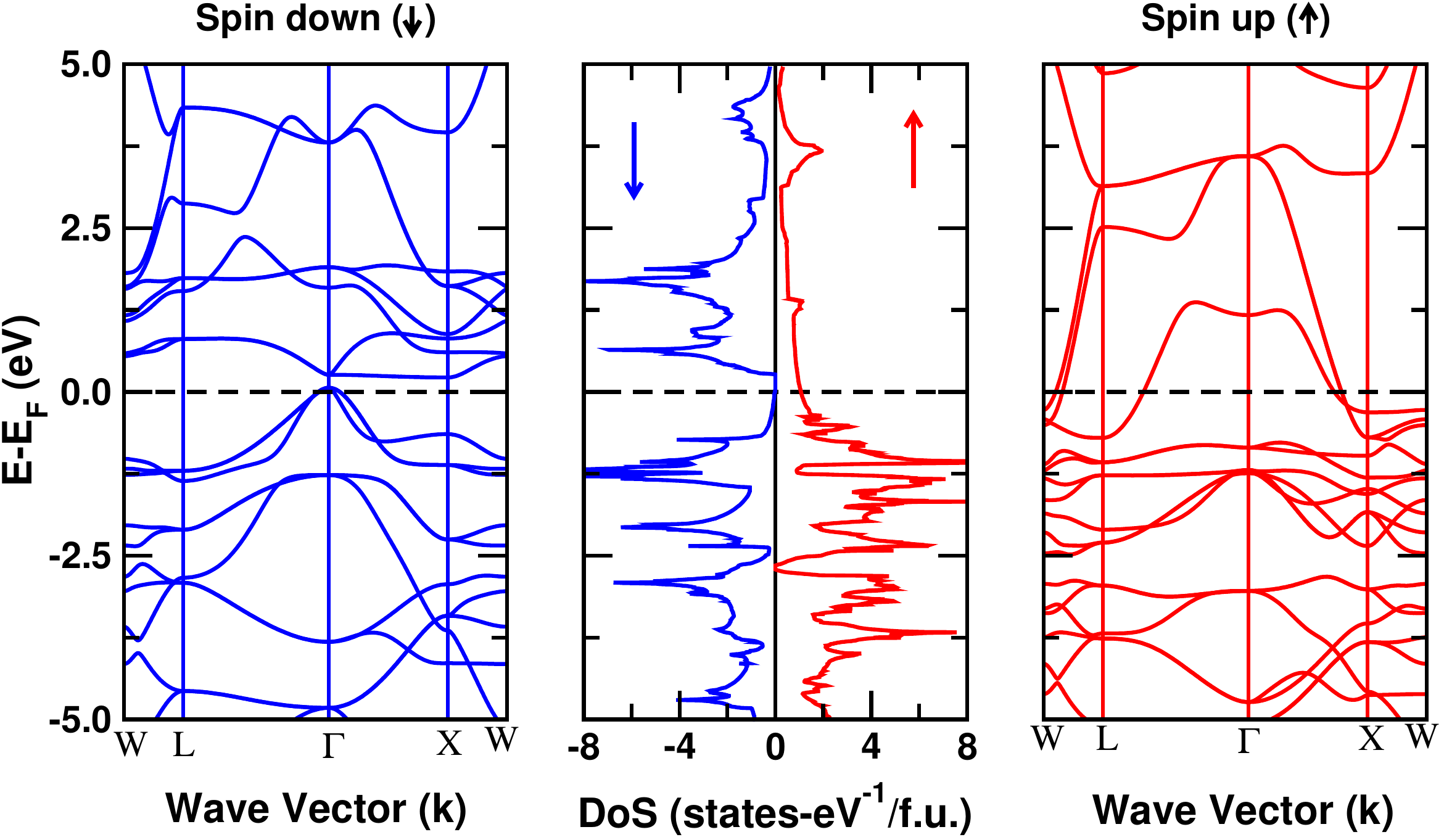}
	\caption{Band Structure and the Density of states (DoS) for CRMG at a$_{rlp}$. The system shows a halfmetallic characteristics having a finite DoS in spin up channel at E$_{F}$ and a non-zero indirect band gap (($\Delta$E$_{\text{g}})_{\downarrow} = 0.23$ eV) in the spin down channel at E$_{F}$. Dashed line represents the Fermi level (E$_F$).}
	\label{fig3}
\end{figure}

\section{Results and Discussion}

\subsection{Structural stability} 
\label{struc-stability}
\par
Quaternary Heusler alloys exhibit LiMgPdSn-type crystal structure (\# $216$) whose primitive cell contains four atoms at the Wyckoff positions, 4a(0, 0, 0), 4b(1/2, 1/2, 1/2), 4c(1/4, 1/4, 1/4) and 4d(3/4, 3/4, 3/4). The preferred occupation depends upon the electronegativities and sizes of individual atoms. Hence, we have calculated the total energies corresponding to different atomic configurations. We have found three non-degenerate atomic configurations, obtained by interchanging constituent atoms at various Wyckoff positions such as Type1, Type2 and Type3 with decreasing stability. The energetics and corresponding configurations are shown in  Fig. {\ref{fig1}}. The most stable configuration (Type1) corresponds to the occupation: Co at 4d, Rh at 4c, Mn at 4b and Ge at 4a Wyckoff sites. On the other hand in Type2 configuration, Co sits at 4b, Rh at 4d, Mn at 4c and Ge at 4a sites. The least stable structure, Type 3, is formed when Co occupies at 4d, Rh at 4b, Mn at 4c and Ge at 4a Wyckoff sites respectively.

\subsection{Electronic structure, magnetic and mechanical properties}
\par
In this section, spin polarized electronic structure, magnetic properties and the mechanical stability of CRMG alloy are presented using {\it{ab-initio}} simulations. 
Figure \ref{fig3} shows spin polarized band structure and density of states calculated at theoretically  relaxed lattice parameter (a$_{rlp} = 5.92 \mathrm{\AA}$). CRMG clearly shows an indirect band gap of $0.23$ eV from $\Gamma \rightarrow X$ in the minority spin channel and finite DoS at E$_{F}$ in the majority spin channel leading to the ideal 100$\%$ spin polarized current in the system. 
Experimental lattice parameter (a$_{elp}$) is measured to be 5.89 $\mathrm{\AA}$ which is not very different from that of a$_{rlp}$. Calculated band structure at a$_{elp}$ looks very similar to that of Fig. \ref{fig3} with a band gap of $0.23$ eV.

The calculated total magnetic moment (m$_{tot}$) is $4.94\ \mu_{B}/f.u.$ at a$_{rlp}$, which follows Slater-Pauling rule, {\cite {PhysRevB.83.184428,PhysRevB.66.174429,Slat1,Paul1}} according to which, $m = (N_v - 24) \mu_B$,
where $N_v$ is the number of valence electrons per unit cell. Our experimentally measured total moment is $4.9\ \mu_{B}/f.u.$.  We have also calculated the Curie temperature (T$_C$) using the model presented in Refs. \onlinecite{0022-3727-40-6-S01, doi:10.1063/1.3062812}. The calculated T$_C$ comes out to be $915$ K ($917$ K) with a$_{elp}$ (a$_{rlp}$), which compares fairly well with those measured by us ($\sim$ $760$ K).

We have also checked the mechanical stability of the alloy by performing lattice dynamics calculations. The linear response parameter e.g. elastic constants (C$_{ij}$) are nothing but the proportionality factor of stress and strain of the crystal under applied force. Calculated values of $C_{ij}$ for CRMG at equilibrium lattice constant (a$_{rlp}$) are C$_{11} = 259.83$, C$_{12} = 167.41$, C$_{44} = 98.00$ N/m$^2$. The condition for mechanical stability of cubic crystals, the so called Born-Huang criteria,{\cite{BHC}} is defined as, 
\[
C_{11} > 0, \; C_{44} > 0, \; C_{11} > C_{12}, \; (C_{11}+2C_{12}) > 0,   
\]
These criteria very well hold in the present case and hence CRMG is mechanically stable. The other elastic moduli such as Bulk modulus (B), shear modulus (G), Young's modulus (Y) and the anisotropy factor (A) are also calculated and are given by $198.22$, $72.83$, $194.65$ and $2.12$ respectively.

\begin{table}[t]
	\begin{ruledtabular}
		\begin{tabular}{|c| c| c| c| c| c|}
			a (\AA)  & Pressure & n$_{\uparrow}$(E$_{F}$)& n$_{\downarrow}$(E$_{F}$) & ($\Delta$E$_{g}$)$_{\downarrow}$
			& Nature \\
			\hline
			5.00 & 327.08 & 1.40 & 2.10 & no-gap & Metallic\\
			\hline
			5.10 & 252.96 & 1.13 & 1.92 & no-gap & Metallic\\
			\hline
			5.20 & 193.39 & 0.92 & 1.86 & no-gap & Metallic\\
			\hline
			5.30 & 145.24 & 0.90 & 1.86 & no-gap & Metallic\\
			\hline
			5.40 & 106.54 & 0.87 & 1.48 & no-gap & Metallic\\
			\hline
			5.50 & 75.32 & 0.87 & 1.04 & no-gap & Metallic\\
			\hline
			5.60 & 50.30 & 0.89 & 0.49 & no-gap & Metallic\\
			\hline
			5.70 & 30.27 & 0.94 & 0.00 & 0.32 & HM\\
			\hline
			5.80 & 14.57 & 1.00 & 0.00 & 0.28 & HM\\
			\hline
			5.85 & 8.09 & 1.03 & 0.00 & 0.26 & HM\\
			\hline
			5.89(a$_{elp}$) & 3.49 & 1.05 & 0.00 & 0.25 & HM\\
			\hline
			5.92(a$_{rlp}$) & 0.00 & 1.07 & 0.00 & 0.23 & HM\\
			\hline
			6.00 & -6.86 & 1.12 & 0.05 & no-gap & Metallic\\
		\end{tabular}
	\end{ruledtabular}
	\caption{Effect of pressure on the electronic properties of CRMG. Upto 5.70\AA\ (pressure of 30.27 GPa), system retains its halfmetallic characteristics with the variable band-gap (($\Delta$E$_{g}$)$_{\downarrow}$) and DoS (n$_{\uparrow}$(E$_{F}$) and n$_{\downarrow}$(E$_{F}$)). A phase transition from halfmetallic (HM) to metallic occurs below 5.70\AA  (i.e. above 30.27 GPa). Band gap, DoS and pressure are measured in eV, states/eV-f.u. and GPa respectively.}
	\label{tab1}
\end{table}

\begin{table}[t]
	\begin{ruledtabular}
		\begin{tabular}{|c| c| c| c| c| c|}
			\hline
			a (\AA)  & m$_{Co}$ & m$_{Rh}$ & m$_{Mn}$
			& m$_{Ge}$ &  m$_{tot}$\\
			\hline
			5.00 & 0.69 & 0.38 & 2.44 & 0.02 & 3.52 \\
			\hline
			5.10 & 0.76 & 0.39 & 2.65 & 0.01 & 3.80 \\
			\hline
			5.20 & 0.87 & 0.42 & 2.81 & 0.01 & 4.12 \\
			\hline
			5.30 & 0.97 & 0.45 & 2.95 & 0.01 & 4.38 \\
			\hline
			5.40 & 1.06 & 0.48 & 3.06 & 0.01 & 4.60 \\
			\hline
			5.50 & 1.14 & 0.49 & 3.15 & 0.00 & 4.77 \\
			\hline
			5.60 & 1.20 & 0.50 & 3.21 & -0.01 & 4.89 \\
			\hline
			5.70 & 1.20 & 0.47 & 3.26 & -0.01 & 4.91 \\
			\hline
			5.80 & 1.20 & 0.44 & 3.31 & -0.02 & 4.92 \\
			\hline
			5.85 & 1.20 & 0.42 & 3.33 & -0.03 & 4.93 \\
			\hline
			5.89(a$_{elp}$) & 1.20 & 0.41 & 3.36 & -0.03 & 4.93 \\
			\hline
			5.92(a$_{rlp}$) & 1.20 & 0.39 & 3.37 & -0.03 & 4.94 \\
			\hline
			6.00 & 1.20 & 0.37 & 3.42 & -0.04 & 4.95 \\
			\hline
		\end{tabular}
	\end{ruledtabular}
	\caption{Pressure effect on the atom projected and total magnetic moment (in Bohr magneton unit) of CRMG. All the constituent transition metal elements are ferromagnetically aligned. Notably, the system retains its ferromagnetic characteristics through out the pressure range i.e. there is no magnetic phase transition.}
	\label{tab2} 
\end{table}

\begin{table*}
	\begin{ruledtabular}
		\begin{tabular}{|c| c| c| c| c| c| c| c| c|}
			\hline
			a (\AA)  & C$_{11}$ & C$_{12}$ & C$_{44}$ & B & G & Y & A & Born-Huang Criteria\\
			\hline
			5.00 & 939.74 & 1043.46 & 480.81 & 1008.88 & -215.75 & -696.94 & -9.27 & $\text{\sffamily X}$\\
			\hline
			5.10 & 821.35 & 899.80 & 418.18 &  873.65 & -113.98 & -357.49 &  -10.66 & $\text{\sffamily X}$\\
			\hline
			5.20 & 695.08 & 754.68 & 360.70 &  734.81 & -60.94 & -188.02 & -12.10 & $\text{\sffamily X}$\\
			\hline
			5.30 & 612.80 & 613.87 & 309.33 & 613.51 & 66.25 & 191.85 &  -578.19 & $\text{\sffamily X}$\\
			\hline
			5.40 & 561.04 & 494.75 & 264.21 & 516.85 & 118.96 & 331.46 & 7.97 & $\checkmark$\\
			\hline
			5.50 & 513.71 & 399.18 & 225.00 & 437.36 & 131.85 & 359.43 & 3.93 & $\checkmark$\\
			\hline
			5.60 & 456.75 & 320.94 & 189.77 & 366.21 & 126.70 & 340.79 & 2.79 & $\checkmark$\\
			\hline
			5.70 & 392.69 & 263.83 & 156.95 & 306.78 & 110.51 & 295.99 & 2.44 & $\checkmark$\\
			\hline
			5.80 & 327.07 & 215.85 & 128.44 & 252.92 & 92.33 & 246.93 & 2.31 & $\checkmark$\\
			\hline
			5.85 & 298.53 & 195.20 & 115.58 & 229.64 & 84.12 & 224.91 & 2.24 & $\checkmark$\\
			\hline
			5.89(a$_{elp}$) & 277.01 & 179.72 & 105.88 & 212.15 & 77.89 & 208.20 & 2.18 & $\checkmark$\\
			\hline
			5.92(a$_{rlp}$) & 259.83 & 167.41 & 98.00 & 198.22 & 72.83 & 194.65 & 2.12 & $\checkmark$\\
			\hline
			6.00 & 224.37 & 142.13 & 81.56 & 169.54 & 62.22 & 166.32 & 1.98 & $\checkmark$ \\
			\hline
		\end{tabular}
	\end{ruledtabular}
	\caption{Effect of pressure on elastic constants (C$_{ij}$), Bulk modulus (B), Shear modulus (G), Young's modulus (Y) and anisotropy factor (A) for CRMG. The alloy is mechanically stable upto $5.40\ \mathrm{\AA}$, satisfying Born-Huang criteria. Any further decrease in $a$ (increase in pressure) leads instability. The symbol, $\checkmark$($\text{\sffamily X}$) represents whether the Born-Huang criteria is satisfied or not. All the elastic parameters are in GPa unit. }
	\label{tab3}
\end{table*}

\subsection{Pressure effect on electronic structure and mechanical stability}
\par

Due to high value of spin polarization, the half-metallic materials are frequently used in spintronic devices in the form of thin films or multilayer. The lattice parameters of pristine bulk materials usually change when the films or multilayer are grown on appropriate substrates.  As such, it is important to check the effect of lattice parameter variation (or pressure) on the electronic structure of bulk material which can give some hints for the surface properties.
 
Effect of hydrostatic pressure on CRMG alloy has been investigated theoretically by reducing the value of lattice constant from its equilibrium value. Pressure effect on electronic structure, magnetism and mechanical properties are summarized in Table~ \ref{tab1}, \ref{tab2} and \ref{tab3} respectively. The alloy retains its halfmetallic characteristics in a limited range of pressure (upto 5.70$\mathrm{\AA}$ which corresponds to 30.27 GPa pressure) with variable band-gap and DoS value in the minority and majority spin channels at E$_{F}$. Below 3.2$\%$ reduction of lattice parameter (i.e. $<$5.70$\mathrm{\AA}$) with respect to a$_{elp}$, DoS in the minority spin channel grows tremendously and resulted in a complete suppression of the band gap and consequently a phase transition from halfmetallic to metallic state (Table I). 

There is no magnetic transition seen in the considered range of pressure and the system retains its ferromagnetic characteristics, but with decreasing magnitude in m$_{tot}$ (Table II). Thus, the system is robust against magnetic transition.

Table III summarizes the effect of pressure on the mechanical properties of CRMG alloy. Elastic constant, C$_{ij}$ and various elastic moduli are affected very significantly. As the pressure increases, the elastic moduli B, G, Y and A reduces.  Upto $5.40\ \mathrm{\AA}$(i.e. $106.5$ GPa), CRMG satisfies the Born-Huang criteria but below $5.40\ \mathrm{\AA}$, the alloy becomes mechanically unstable.

\begin{figure}[b]
	\centering
	\includegraphics[width=0.9\linewidth]{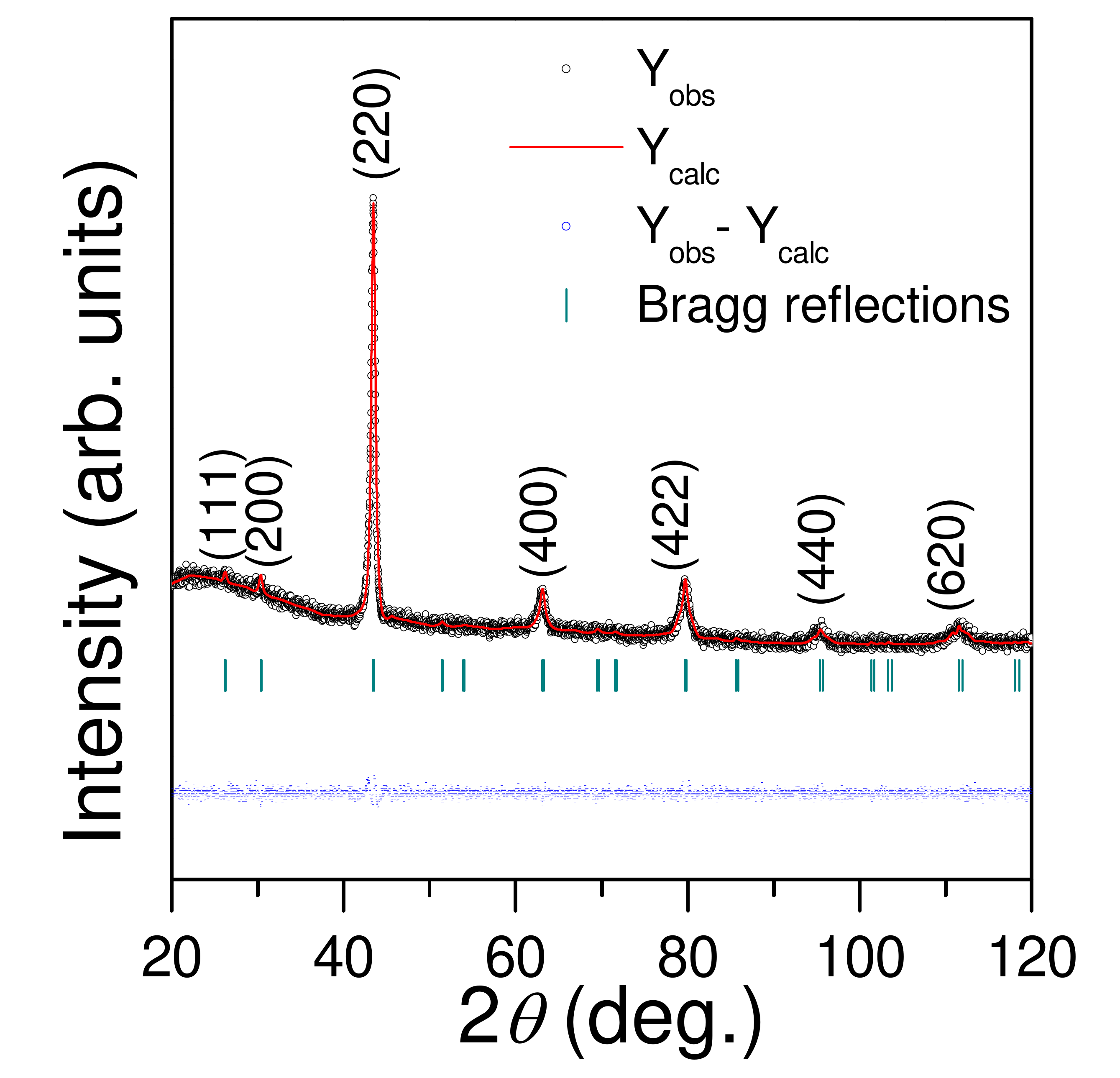}
	\caption{Rietveld refined XRD pattern of CRMG alloy. $\mathrm{Y_{obs}}$ and $\mathrm{Y_{calc}}$ are the observed and calculated scattering intensities. }
	\label{XRD}
\end{figure}

\begin{figure}[b]
	\centering
	\includegraphics[width=0.8\linewidth]{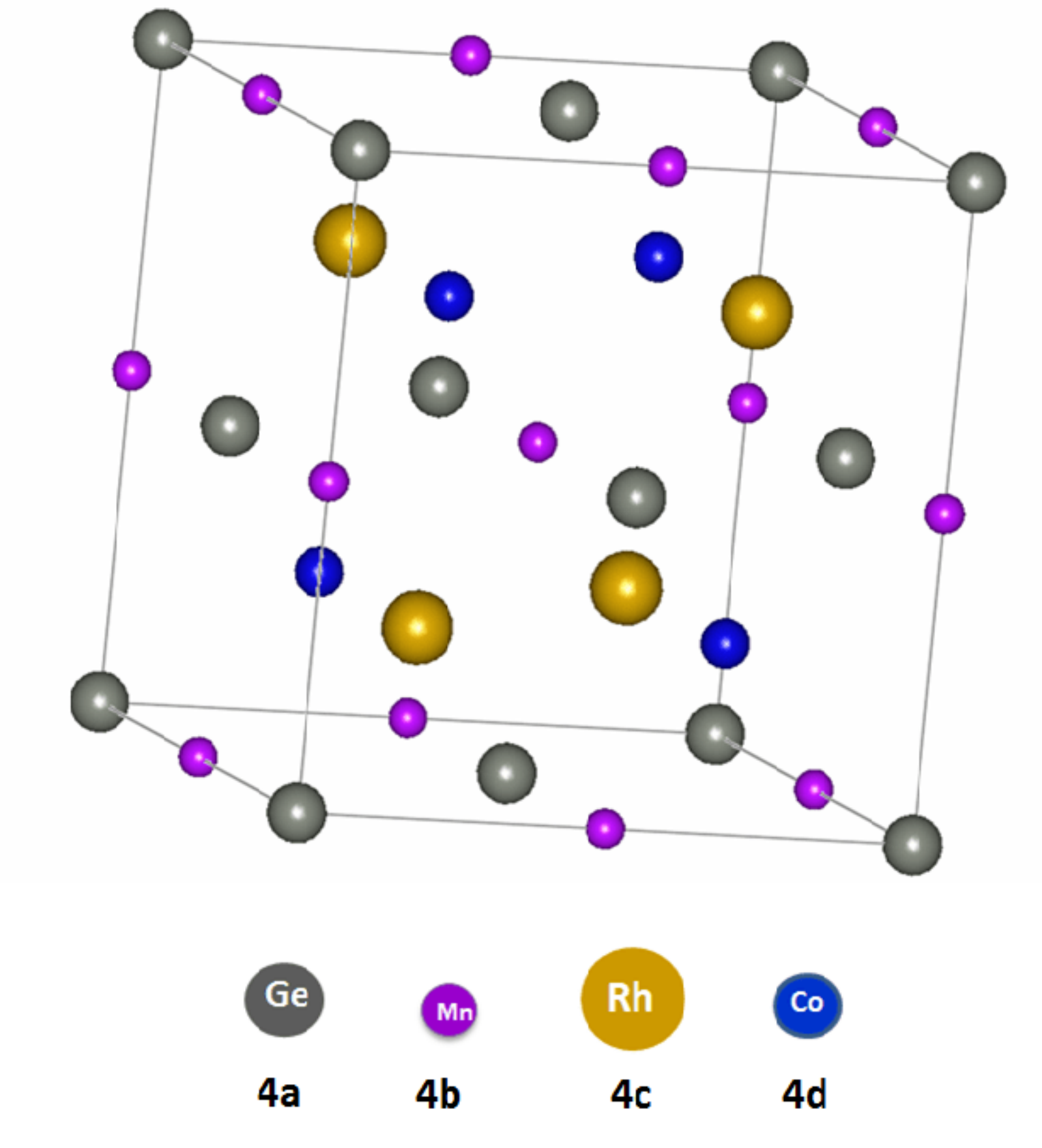}
	\caption{Measured Crystal Structure of CRMG alloy }
	\label{fig5}
\end{figure}

\subsection{X-ray diffraction}
\begin{figure}[t]
\centering
\includegraphics[width=1.0\linewidth]{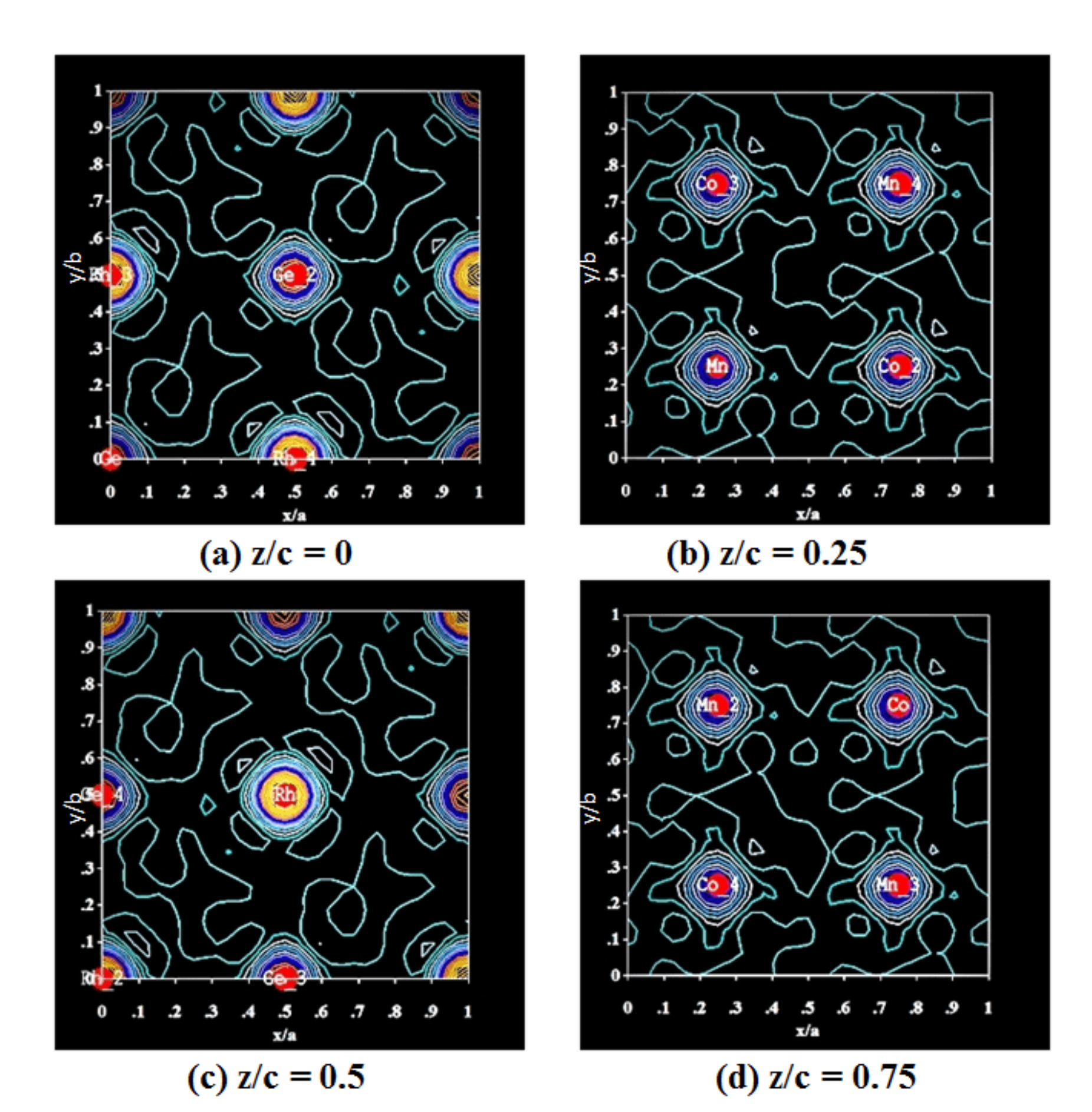}
\caption{Electronic density of individual atoms in the unit cell of CRMG alloy at (a) z/c = 0, (b) z/c = 0.25, (c) z/c = 0.5 and (d) z/c = 0.75}
\label{fig:CRMGElectrondensityfinal}
\end{figure}

 Figure {\ref{XRD}} shows the Rietveld refinement of the room temperature XRD pattern using Fullprof suite. It is clear from the pattern that the alloy CRMG exhibits a cubic structure. The quaternary Heusler alloys exhibit LiMgPdSn-type structure whose primitive cell contains four atoms on the wyckoff positions 4a, 4b, 4c and 4d. Out of various possibilities for the occupancy of four atoms at different sites, only three atomic configurations turn out to be energetically distinct for this structure.\cite{PhysRevB.92.224413,0953} Due to the underlying symmetry of $F\bar{4}3m$ (\#216) space group, interchange of atoms between 4a and 4b or 4c and 4d Wyckoff positions does not change the overall symmetry and hence the total energy of the compound. The presence of super-lattice reflections  in the XRD pattern confirms CRMG alloy to crystallize in cubic Heusler structure. The lattice parameter was found to be 5.89\ $\mathrm{\AA}$, which compares fairly well with theory (a$_{rlp}=5.92~ \mathrm{\AA}$). The best fit of the observed intensities was obtained when Co, Mn, Rh and Ge atoms were assigned the Wyckoff positions 4d, 4b, 4c and 4a respectively, which corresponds to a Type1 structure (Fig. \ref{fig1}). The measured crystal structure of CRMG alloy is shown in Fig. \ref{fig5}. The super-lattice reflections are not very intense, which can be due to some disorder in the system. In CRMG, X(Co), Y(Mn) and Z(Ge) atoms are from the same period and their atomic scattering factors are nearly identical. In such cases, it is difficult to find out the extent of disorder using XRD. EXAFS is helpful in determining the short range chemical environment around the atoms. Thus, to further probe the local surroundings of Co, Ge and Mn site, EXAFS  measurements have been performed on CRMG alloy, as discussed later.
The electronic densities of the constituent elements at different vertical cut are shown in Fig. \ref{fig:CRMGElectrondensityfinal}(a-d), which is generated from the XRD refinement using GFourier program in Fullprof suite. Figure \ref{fig:CRMGElectrondensityfinal}(c) and Fig. \ref{fig:CRMGElectrondensityfinal}(d) clearly suggest that most of the charge is distributed around Rh site. Co and Mn are surrounded by intermediate charge while Ge has a charge density which is less than that of Rh but greater than that of Co and Mn atomic sites. Thus the position of heavier element in the unit cell is reflected by the most dense electron density contour.

\begin{figure}[t]
	\centering
	\includegraphics[width=0.9\linewidth]{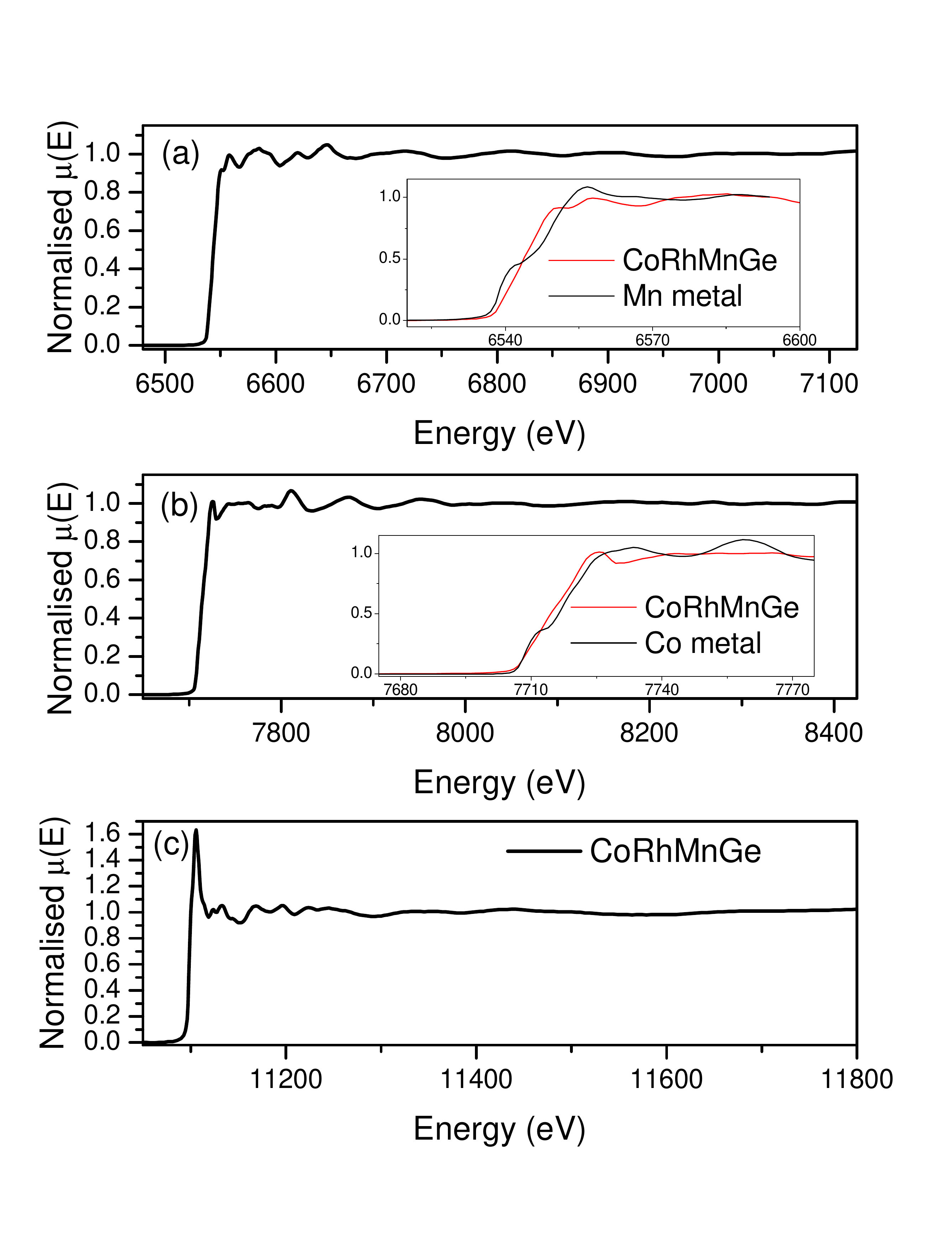}
	\caption{Normalized EXAFS spectra of (a) Co, (b) Mn and (c) Ge of CRMG sample. The insets of (a) and (b) show the XANES spectrum of CRMG at Co and Mn K-edge (red line) along with the corresponding metal foil reference (black line).}
	\label{fig:EXAFS1}
\end{figure}

\subsection{EXAFS Analysis}
\begin{figure}
\centering
\includegraphics[width=1.0\linewidth]{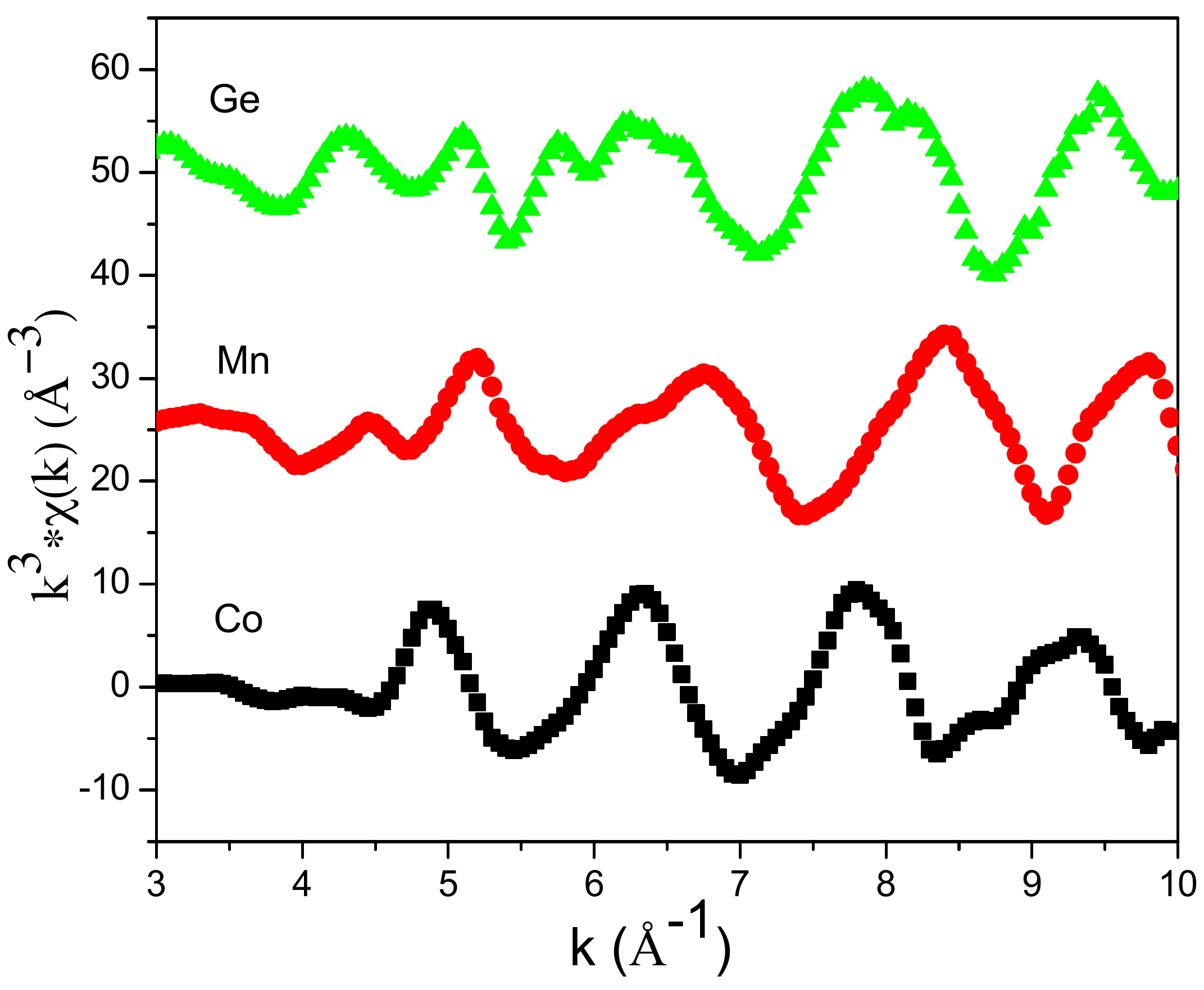}
\caption{$k^3*\chi(K)$ vs. k spectra for Co, Mn and Ge of k- edge of CRMG.}
\label{fig:EXAFS2}
\end{figure}

Figure \ref{fig:EXAFS1} shows the normalized EXAFS spectra of (a) Co (b) Mn and (c) Ge for CRMG alloy. The insets of (a) and (b) show the XANES (X-ray absorption near edge structures) spectrum of CRMG at Co and Mn K-edge (red line). We have checked the possibility of oxidation of the samples by comparing the XANES spectra of the samples with that of standard metal foils and commercial powders of respective oxides. The spectra of pure metal foil reference is shown by black line in the insets. We found that the edge positions and the XANES spectra of CRMG sample  agree fairly well with those of the pure metal foils and do not resemble that of the oxides, thus ruling out the presence of oxides. In order to take care of the oscillations in the absorption spectra, the absorption function $\chi(E)$ is calculated as follows \cite{Dc}
\begin{equation}
\chi(E) = \frac{\mu(E)-\mu(E_0)}{\Delta{\mu(E_0)}}
\end{equation}
Here $E_0$ is absorption edge energy, $\Delta{\mu(E_0)}$ is the bare atom background and $\mu(E)$ is the step in   value at the absorption edge. The $k$-dependent absorption coefficient $\chi(k)$ is then calculated using the relation, 
\begin{equation}
k = \sqrt{\frac{2m(E-E_0)}{\hbar^2}}
\end{equation}
where, m
is the electronic mass. $\chi(k)$ is then weighted by $k^3$ to amplify the oscillation at high k. The $k^3*\chi(k)$  spectra for Co, Mn and and Ge k-edge  are shown in Fig. {\ref{fig:EXAFS2}}. The $k^3 \chi(k)$ functions are also Fourier transformed in the R-space to generate the 
$\chi(R)$ spectra in terms of the real distances from the center of the absorbing atom.
The set of EXAFS data analysis programme available within IFEFFIT software package has been used for final data analysis.\cite{NEWVILLE1995154} This includes background reduction and Fourier transform to derive the  $\chi(R)$ versus R
spectra from the absorption spectra (using ATHENA software), generation of the theoretical EXAFS spectra starting from an assumed crystallographic structure and finally fitting of experimental data with the theoretical spectra using ARTEMIS software.

\begin{figure}[t]
	\centering
	\includegraphics[width=1.0\linewidth]{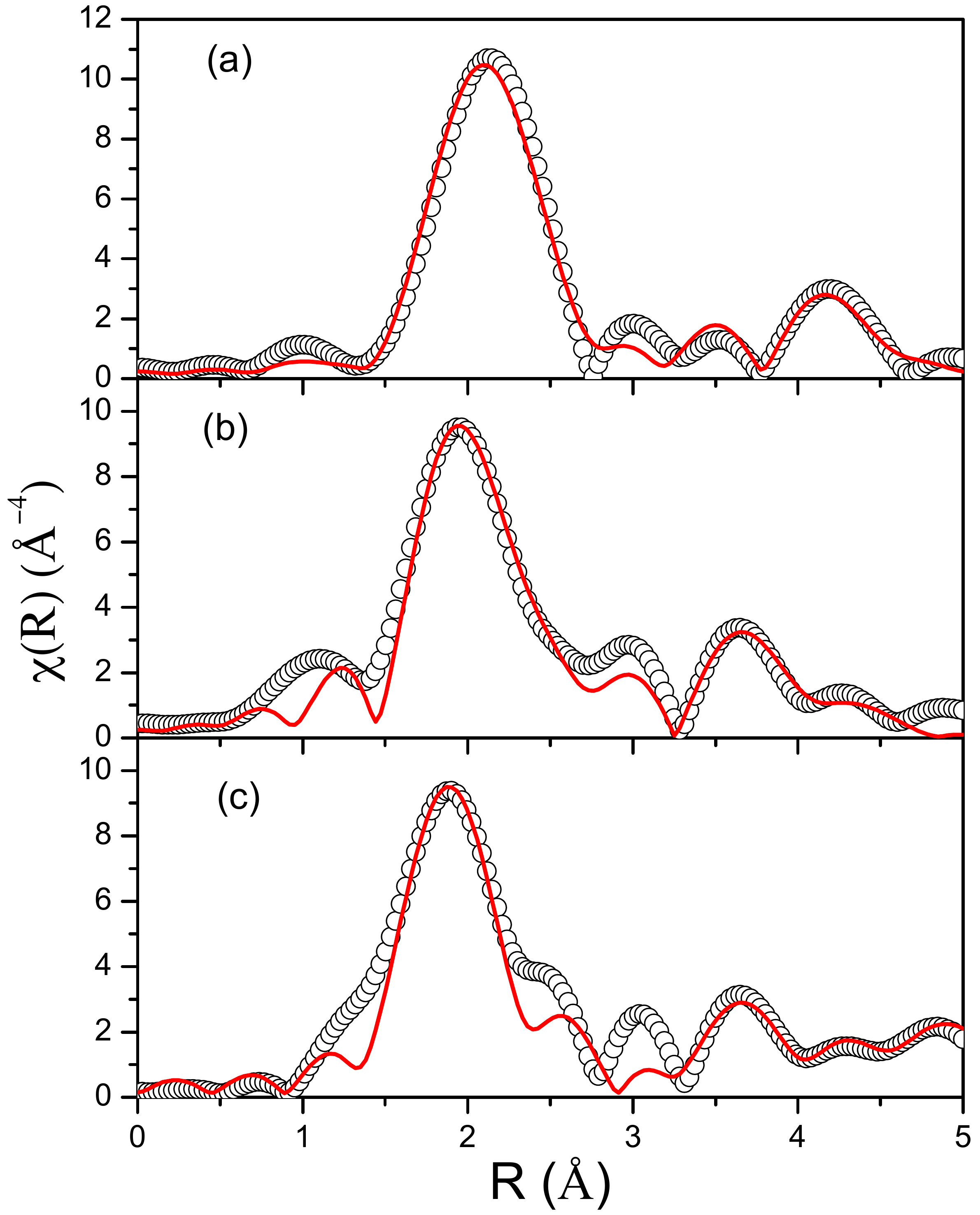}
	\caption{Fourier transformed EXAFS spectra $\chi(R)$ of CRMG at  (a) Co, (b) Mn and (c) Ge K-edge (scatter points) and theoretical fit (red solid line).}
	\label{fig:EXAFS3}
\end{figure}

The $\chi(R)$
versus R spectra generated (Fourier transform range is $k=3.0-10.0$ $\mathrm{\AA^{-1}}$ for the CRMG sample) using the methodology described above are shown in Fig. \ref{fig:EXAFS3}
for CRMG measured at Co, Mn and Ge K-edge. The fitting strategy adopted here was to simultaneously fit multiple data set with multiple edges.\cite{Bainsla2015509,PhysRevB.65.184431} Co, Mn and Ge K edges are fitted simultaneously with common fitting parameters. This fitting reduces the number of independent parameters below the Nyquist criteria and enhances the statistical significance of the fitting model. The goodness of fit has been determined by the value of the $R_{factor}$ defined by:
\small
\begin{equation}
R_{factor}=\sum_{i} {\frac {{[\Im m(\chi_{dat}(r_i)-\chi_{th}(r_i)]^2}+{{[Re(\chi_{dat}(r_i)-\chi_{th}(r_i)}]^2}}{{[\Im m(\chi_{dat}(r_i))]^2}+{[Re(\chi_{dat}(r_i))]^2}}}
\end{equation}
\normalsize
where, $\chi_{dat}$ and $\chi_{th}$ refer to the experimental and theoretical $\chi(r)$ values respectively and $Re$ and $\Im m$ refer to the real and imaginary parts of the respective quantities. In addition to the single scattering paths, multiple scattering paths are also used for the fitting purpose.

The structural parameters (atomic coordination and lattice parameters) of CRMG used for simulation of theoretical EXAFS spectra of the sample have been obtained from XRD results. The best fitted $\chi(R)$ versus $R$
spectra (fitting range R=1.2-5.0 $\mathrm \AA$) of the sample are shown in Fig. \ref{fig:EXAFS3}
along with the experimental data for the measurements carried out at Co, Mn and Ge K-edges. The bond distances, co-ordination numbers (including scattering amplitudes) and disorder (Debye-Waller) factors $(\sigma^2)$, which give the mean square fluctuations in the distances, have been used as fitting parameters. All the three spectra are fitted without site anti-site disorder. Thus, the structural characterization of the CRMG performed using EXAFS analysis reveals a well ordered structure for CRMG (no antisite disorder). All the relevant fitting parameters are shown in Table \ref{tab4}.
\begin{table*}
	\centering
	\caption{Bond length(R), coordination number(N) and Debye-Waller or disorder factor ($\sigma^2$)  obtain by EXAFS fitting for CRMG at Co, Mn and Ge K edge}
	
	\begin{tabular}{|c|c|c|c|c|c|c|c|c|c|c|c|}
		\hline
		\multicolumn{4}{|c|}{\textbf{Co edge}}&
		\multicolumn{4}{c}{\textbf{Mn edge}}&	
		\multicolumn{4}{|c|}{\textbf{Ge edge}} \\
		\hline \rule[-2.5ex]{0pt}{5.5ex}  \hspace*{.3cm}{\textbf{Path}} \hspace*{0.17cm}&\hspace*{.17cm}{\textbf{R(\AA)}}\hspace*{0.17cm}&\hspace*{.17cm}{\textbf{N}}\hspace*{0.17cm}&\hspace*{.17cm}{$\boldsymbol\sigma^2$}\hspace*{0.17cm}&\hspace*{.17cm}{\textbf{Path}}\hspace*{0.17cm}&\hspace*{.17cm}{\textbf{R(\AA)}}\hspace*{0.17cm}&\hspace*{.17cm}{\textbf{N}}\hspace*{0.17cm}&\hspace*{.17cm}{$\boldsymbol\sigma^2$}\hspace*{0.17cm}&\hspace*{.17cm}{\textbf{Path}}\hspace*{0.17cm}&\hspace*{.17cm}{\textbf{R(\AA)}}\hspace*{0.17cm}&\hspace*{.17cm}{\textbf{N}}\hspace*{0.17cm}&{$\boldsymbol\sigma^2$}  \\ 
		\hline \rule[-2ex]{0pt}{5.5ex} $\boldsymbol{\mathrm{Co-Ge}}$  &$2.40\pm0.01$  &4  &$0.0085\pm0.001$  &$\boldsymbol{\mathrm{Mn-Ge}}$  &$2.40\pm0.01$  &4  &$0.001\pm0.0008$  &$\boldsymbol{\mathrm{Ge-Co}}$  &$2.39\pm0.01$  &4  &$0.02\pm0.002$  \\ 
		\hline \rule[-2ex]{0pt}{5.5ex} $\boldsymbol{\mathrm{Co-Rh}}$  &$2.40\pm0.01$  &4&  $0.035\pm0.002$&  $\boldsymbol{\mathrm{Mn-Rh}}$&  $2.40\pm0.01$&  4&  $0.001\pm0.0008$&  $\boldsymbol{\mathrm{Ge-Mn}}$&  $2.39\pm0.01$&  4&  $0.0057\pm0.001$  \\ 
		\hline \rule[-2ex]{0pt}{5.5ex} $\boldsymbol{\mathrm{Co-Mn}}$  &$2.92\pm0.01$  &6  &$0.0142\pm0.001$  &$\boldsymbol{\mathrm{Mn-Co}}$  &$2.89\pm0.02$  &6  &$0.0167\pm0.001$  &$\boldsymbol{\mathrm{Ge-Rh}}$  &$2.92\pm0.01$  &6  &$0.0199\pm0.0008$  \\ 
		\hline \rule[-2ex]{0pt}{5.5ex} $\boldsymbol{\mathrm{Co-Co}}$  &$4.05\pm0.02$  &12  &$0.0255\pm0.0009$  &$\boldsymbol{\mathrm{Mn-Mn}}$  &$4.05\pm0.01$  &12  &$0.0321\pm0.001$  &$\boldsymbol{\mathrm{Ge-Ge}}$  &$4.11\pm0.03$  &12  &$0.0128\pm0.006$  \\ 
		\hline \rule[-2ex]{0pt}{5.5ex} $\boldsymbol{\mathrm{Co-Ge}}$  &$4.60\pm0.03$  &12  &$0.0098\pm0.001$  &$\boldsymbol{\mathrm{Mn-Ge}}$  &$4.58\pm0.02$  &12  &$0.004\pm0.001$  &$\boldsymbol{\mathrm{Ge-Co}}$  &$4.89\pm0.05$  &12  &$0.0039\pm0.001$  \\ 
		\hline \rule[-2ex]{0pt}{5.5ex} $\boldsymbol{\mathrm{Co-Rh}}$  &$4.60\pm0.03$&  12&  $0.0077\pm0.0007$&  $\boldsymbol{\mathrm{Mn-Rh}}$&  $4.63\pm0.02$&  12&  $0.0043\pm0.001$&  $\boldsymbol{\mathrm{Ge-Mn}}$&  $4.89\pm0.05$&  12&  $0.0026\pm0.001$  \\ 
		\hline 
	\end{tabular} 
	\label{tab4}
\end{table*}

\begin{figure}[b]
	\centering
	\includegraphics[width=1.0\linewidth]{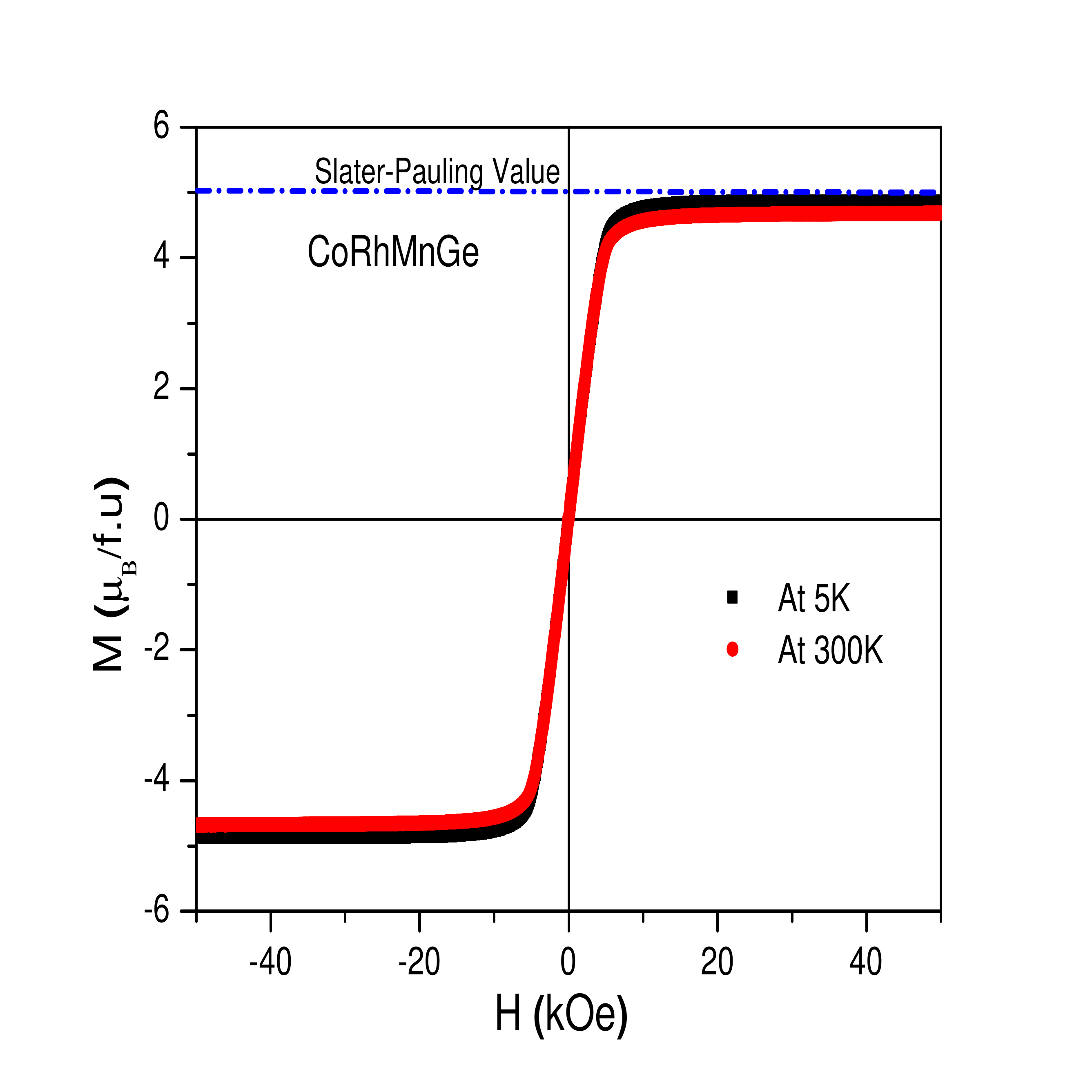}
	\caption{Magnetization curves for CRMG alloy at 5 K  and 300 K.}
	\label{fig:CRMG_MH}
\end{figure}
In the Fourier transformed EXAFS spectrum of Co K-edge (Fig. \ref{fig:EXAFS3}(a)), the main peak near 2.1 $\mathrm{\AA}$ has contribution from Co-Ge (2.40 $\mathrm{\AA}$), Co-Rh (2.40 $\mathrm{\AA}$) and Co-Mn (2.92 $\mathrm{\AA}$) paths. The small peak around 3.5 $\mathrm{\AA}$ at Co K-edge spectrum is the contribution of Co-Co (4.05 $\mathrm{\AA}$) coordination and a small contribution of higher bond length paths. The third peak at 4.25 $\mathrm{\AA}$ is the contribution from Co-Ge (4.60 $\mathrm{\AA}$) and Co-Rh (4.60 $\mathrm{\AA}$) coordination shells with multiple scattering paths. The use of multiple scattering paths does not include any further independent parameters as all the multiple scattering parameters are defined in terms of single scattering path parameters.\cite{PhysRevB.65.184431} In case of Ge K-edge (Fig. \ref{fig:EXAFS3}c), the first peak has the contribution from Ge-Co (2.39 $\mathrm{\AA}$), Ge-Mn (2.39 $\mathrm{\AA}$) and Ge-Rh (2.92 $\mathrm{\AA}$) paths. The peak at 3 $\mathrm{\AA}$ could not be fitted well as there is no coordination shell  present at this distance. The peak at 3.6 $\mathrm{\AA}$ is the contribution from Ge-Ge (4.11 $\mathrm{\AA}$) path. The double peak  between 4 $\mathrm{\AA}$ and 5 $\mathrm{\AA}$ is fitted with Ge-Co (4.89 $\mathrm{\AA}$), Ge-Mn (4.89 $\mathrm{\AA}$) and multiple scattering paths. Similarly, in case of Mn K-edge (Fig. \ref{fig:EXAFS3}(b)) the first peak has the contribution from Mn-Ge (2.40 $\mathrm{\AA}$), Mn-Rh (2.40 $\mathrm{\AA}$) and Mn-Co (2.89 $\mathrm{\AA}$). The region between 3.5 $\mathrm{\AA}$ and 4.5 $\mathrm{\AA}$ is fitted with Mn-Mn, Mn-Ge and Mn-Rh coordination shells with multiple scattering paths.

\subsection{Magnetic Properties}

\begin{figure}[b]
	\centering
	\includegraphics[width=1.0\linewidth]{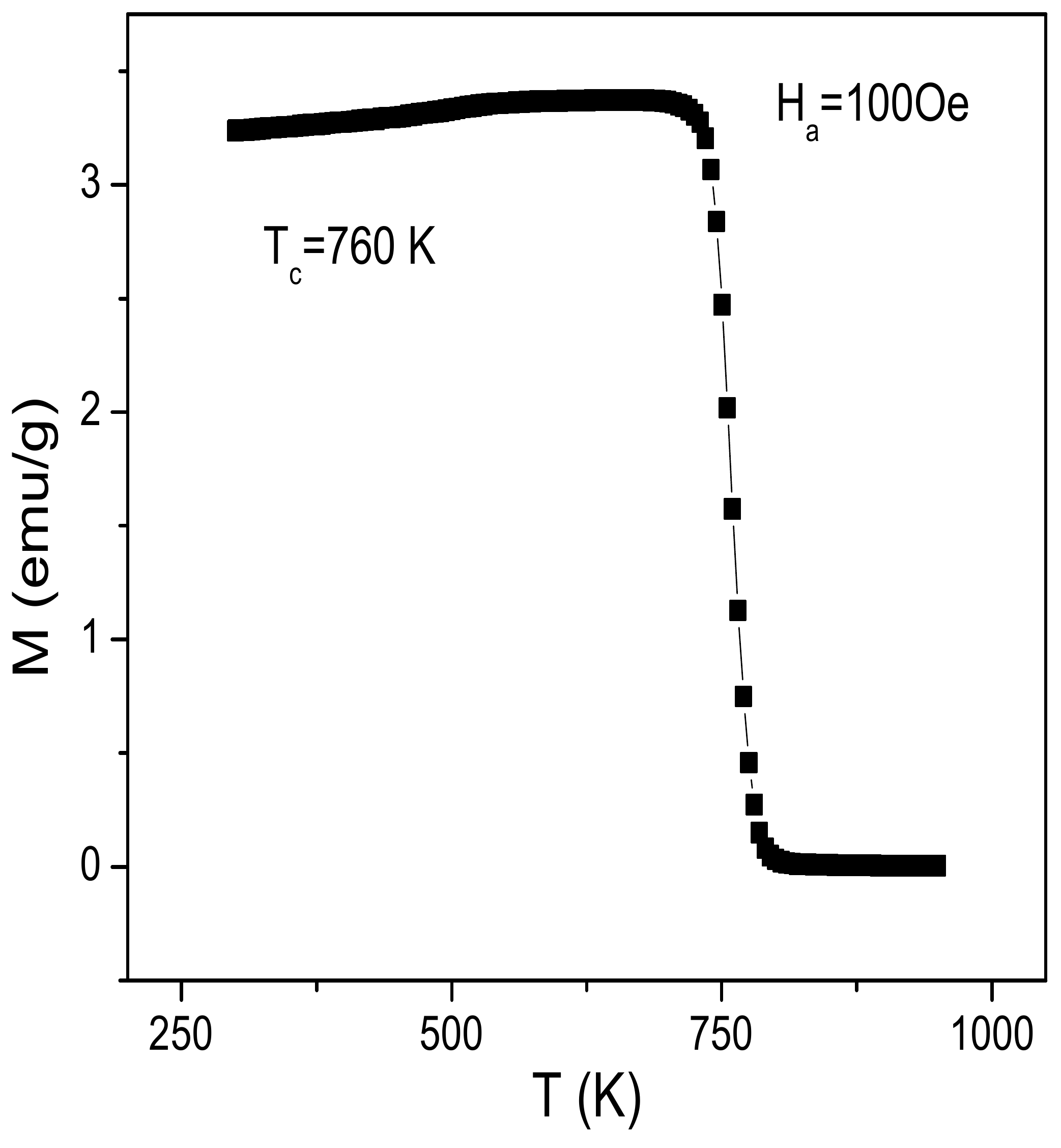}
	\caption{Temperature dependence of magnetization at 100 Oe. $T_c$ is calculated from the minima of the first order derivative of $M$ vs. $T$ curve.}
	\label{fig:CRMGMT}
\end{figure}

\begin{figure*}[t]
	\centering
	\includegraphics[width=0.9\linewidth]{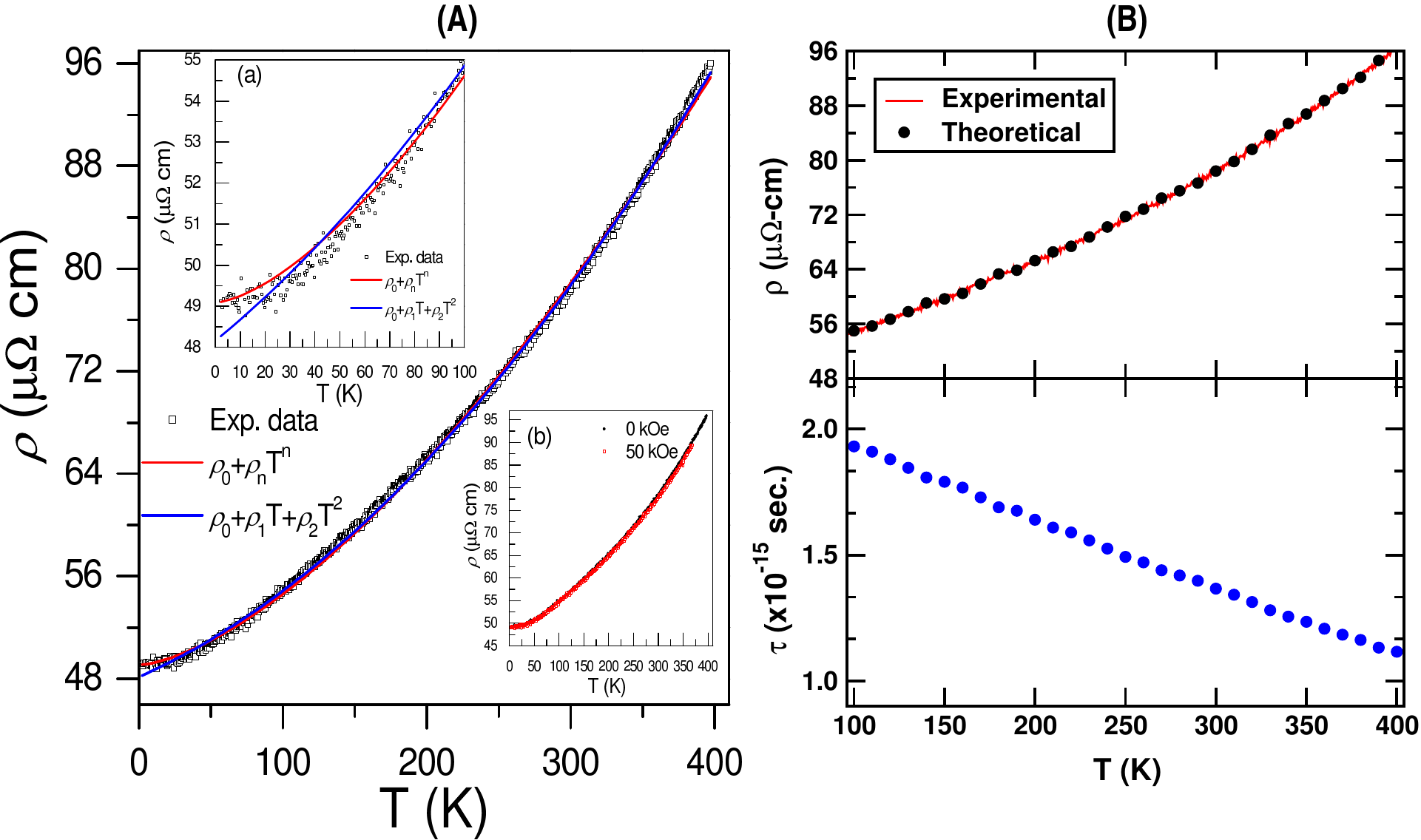}
	\caption{(A) Temperature dependence of electrical resistivity($\rho$) for CRMG at zero field. The raw data is fitted with two functions, a general power law and a quadratic fit. Inset (a) shows a closer look of the $\rho$ curve between $0-100$ K. Inset (b) shows $\rho$ vs. T at $0$ and $50$ kOe. (B) (top) Comparison of measured $\rho$ (red line) with those simulated (solid circle). Temperature dependence of the carrier relaxation time ($\tau$) (bottom).  }
	\label{fig:Res-fits-final}
\end{figure*}

\begin{table*}
	\centering
	\caption{Fitting parameters for zero field electrical resistivity vs. $T$ for CRMG alloy. Two different fitting procedure involve a general power law and a quadratic fit. Residual resistivity ($\rho_0$) in both the case is same i.e. $\rho_0=48.16\ \mu\Omega$.}
	
	\begin{tabular}{|c|c|c|c|}
		\hline & & \\
		$\rho(T) = \rho_0 + \rho_n\:T^n$ &
		$\rho_n = 0.0047(1)$$\mu\Omega \mathrm{\:cm\:K^{-n}}$ & $n=1.535(4)$ \\
		\hline  & &\\
		$\rho(T) = \rho_0+ \rho_1 T + \rho_2 T^2$ & $\rho_1 = 0.0496(4)$ $\mu\Omega \mathrm{\:cm\:K^{-1}}$ & $ \rho_2 = [1.742(1)]\times{10^{-4}}$$\mu\Omega\:\mathrm{cm\:K^{-2}}$ \\
		\hline
	\end{tabular} 
	\label{tab5} 
\end{table*}

Figure \ref{fig:CRMG_MH} shows the isothermal magnetization curves for the CRMG alloy at 5 K and 300 K. The absence of hysteresis shows the soft magnetic nature of the alloy. The saturation magnetization values ($M_s$) at 5 K and 300 K were found to be 4.9 $\mu_B/f.u.$ and 4.7 $\mu_B/f.u.$ respectively. The obtained $M_s$ value (at 5 K) is in close agreement with those expected from Slater-Pauling rule (5 $\mu_B/f.u.$ ) which gives a strong indication of the half-metallic nature of this alloy.

Figure \ref{fig:CRMGMT} shows the temperature dependence of the magnetization at a constant field of 100 Oe. The Curie temperature ($T_c$) is found to be $\sim760$ K.

\subsection{Transport properties}
Figure \ref{fig:Res-fits-final}(A) shows the measured temperature dependence of resistivity for CRMG at $0$ kOe.
The inset (b) of Fig. \ref{fig:Res-fits-final}(A) presents the $\rho$ versus T at two different fields $0$ kOe and $50$ kOe, indicating insignificant dependence of $\rho$ on the field. To closely investigate the half-metallic nature of CRMG, we have used two different functions for fitting the zero field resistivity curve. The residual resistivity is found to be $48.16$ $\mu\Omega$ cm. In the first approach, we have fitted the resistivity data to a general power law,
\[
\rho(T) = \rho_0 + \rho_n T^n 
\]
where $\rho_0$ is the residual resistivity. The value of n, when fitted in temperature range 2-400 K is found to be 1.53. 

In the second approach, the resistivity curve was fitted using the equation
\begin{equation}
\rho(T) = \rho_0+ \rho_{phonon} + \rho_{magnon} = \rho_0+ \rho_1 T + \rho_2 T^2
\end{equation}
where $\rho_1$ and  $\rho_2$ are arbitrary constants, $\rho_0$ is the residual resistivity which originates from the scattering of conduction electrons by the lattice defects, impurities etc., $\rho_{phonon}$  and $\rho_{magnon}$ arise due to scattering of phonons and magnons respectively.
With the quadratic temperature dependence, the curve fits well for temperature $\mathrm{T} > 35$ K as shown in the inset (a) of Fig. \ref{fig:Res-fits-final}(A). At low temperatures, however, the electrical resistivity is found to be almost independent of temperature (2-35 K). Similar behavior was also observed for other half-metallic ferromagnets like CoRuFeSi\cite{Bainsla2015631} and $\mathrm{Co_2FeSi}$ \cite{PhysRevLett.110.066601}. The results of the fitting are given in Table \ref{tab5}. It can be seen that the dominant contribution is from the electron-phonon scattering (T-dependence). 

In a half-metallic ferromagnet, the states at E$_{F}$ are completely spin-polarized, and hence spin-flip scattering is not possible\cite{1989JPCM1.2351O,Kubo-1972,1989JPCM1.2351O} due to absence of electrons in the minority band gap at E$_{F}$. Thus, for a half-metallic ferromagnet,  $T^2$ term is expected to be absent in the resistivity. The insignificant contribution of  $T^2$ term in the resistivity data for CRMG alloy, indirectly supports the half-metallic nature.

Figure \ref{fig:Res-fits-final}(B)(top) shows a comparison of  our theoretically simulated resistivity (solid circle) with those measured (red line).  Within Boltztrap method, the resistivity (or the electrical conductivity) is calculated in units of the relaxation time ($\tau$). The bottom panel of Fig. \ref{fig:Res-fits-final}(B) shows the temperature dependence of these relaxation time, which varies in the range $1-2$ femtoseconds and hence lie in the typical range of $\tau$ as for most other standard  compounds. The comparison of resistivity ($\rho$) is shown only between $100-400$ K, because Boltztrap method is based on a semiclassical theory and it relatively yields more accurate results at higher temperatures.  This is mainly because for low temperatures, fewer bands are included in the summation for resistivity expression, and thus a much higher k-point mesh is necessary to obtain accurate estimations of the electrical properties.

\section{Summary and Conclusion}
In summary, we have studied the structural, electronic, magnetic and transport properties of equiatomic quaternary Heusler alloy, CoRhMnGe by means of both theory and experiment. Experimentally prepared alloy crystallizes in Y-type structure with almost no signature of intrinsic disorder. Ab-initio simulation confirms the stability of the measured crystal structure. It also confirms the half-metallic nature of the alloy (leading to high spin polarization) which is supported by the experimentally observed (almost) integer moment ($\sim$ 5.0 $\mu_B$) at 5K. In addition, measured electrical resistivity also indirectly supports the half-metallic behavior in the alloy. Electrical resistivity is also calculated theoretically, which compares fairly well with experiment. Carrier relaxation time ($\tau$) is shown to lie between $1-2$ fs, which is the usual range of $\tau$ in most compounds. Pressure studies reveal that half metallic nature of CRMG remains robust within a limited range of pressure (upto 30.27 GPa) beyond which it becomes metallic. There is, however, no evidence of magnetic phase transition. Electronic energy and lattice dynamics calculations show that the system is chemically as well as mechanically stable. Summarizing all these properties  along with high spin polarization and relatively large value of T$_{C}$ ($\sim760$K) makes CRMG a potential candidate for spintronic applications.

\section*{Acknowledgments}
Deepika Rani would like to thank Council of Scientific and Industrial Research (CSIR), India for providing Junior Research Fellowship. Enamullah (an Institute Post Doctoral Fellow) acknowledges Indian Institute of Technology Bombay for necessary funding to support this research. AA acknowledges DST-SERB (SB/FTP/PS-153/2013) for funding
to support this research.

\bibliography{bib}

\end{document}